\documentclass[global]{svjour}
\usepackage{amssymb}
\usepackage{amsfonts}
\usepackage{amsmath}

\input{tcilatex}
\begin{document}

\title{Second order photon loops at finite temperature and charge
renormalization}
\author{Samina S. Masood \inst{1} and Mahnaz Q. Haseeb \inst{2}}
\institute{Department of Physics, University of Houston Clear Lake, \\
Houston TX 77058 \\
smasood@uhcl.edu\\
Department of Physics, COMSATS Institute of Information Technology,
Islamabad, Pakistan \\
mahnazhaseeb@comsats.edu.pk }
\maketitle

\begin{abstract}
We present two loop corrections to photon self energy at finite temperature
in real time formalism. An expression for renormalized coupling constant has
been derived in a form that is relevant for all temperature ranges of
interest in QED, specifically for temperatures around T $\sim $ m, where $m$%
\ is electron mass. Temperature dependence is mainly contributed by hot
fermions at $T$ $\geq m$. We use the calculations of vacuum polarization to
determine the dynamically generated mass of photon, Debye screening length,
plasma frequency up to order $\alpha ^{2}$\ as well as the electromagnetic
properties of a medium at $m\leq T\leq 2m$\ temperature. For higher
temperatures, the existing renormalization scheme does not work well because
of the increase in coupling constant. To exactly determine the validity of
renormalization scheme, higher order calculations are required. The
temperature{\small \ T }$\sim $ m is of specific interest from the point of
view of the early universe. Such calculations have acquired more
significance recently due to the possibility of producing $e^{+}e^{-}$\
plasmas in laboratory.

{\small \ }PACS numbers: 11.10.Wx, 12.20.-m, 11.10.Gh, 14.60.Cd
\end{abstract}

\keywords{}

\section{Introduction}

Finite temperature effects [1-19] are now extensively used not only in
particle physics but also to study relativistic plasmas. The possibilities
to create ultra-relativistic electron positron plasmas with high-intensity
lasers has generated renewed interest in finite temperature QED to study
processes of pair-production [20-23] in QED plasmas. Lasers pulses hitting a
thin gold foil can heat up the electrons in the foil up to several MeVs
leading to pair creation [24-30]. The latest laser facilities plan to attain
colossal intensities of $10^{26}W/cm^{2}$ which mark a substantial
improvement over the ones highest recorded ($10^{22}W/cm^{2}$), at Hercules
facility in Michigan [31]. This high temperature plasma might have been
present in the very early universe, right after the lepton production
started. But, the density effects were still ignorable at this epoch in the
early universe.

Quantum Field Theory (QFT) assumes particles to be analogous to excitations
of a harmonically oscillating field which permeates space-time. Modern
interpretation of vacuum describes it as an absence of particles, but not
devoid of energy and fields. Vacuum can be treated as a hypothetical bath of
virtual hot particles which can virtually mediate interactions between real
particles for un-measureable short intervals of time. Virtual
electron-positron pairs couple to photons through loops and give photon a
temperature dependent finite size, charge and mass. Particles propagating in
vacuum are assumed to be those with interactions switched off. The exact
state of all the background particles is unknown when these particles
propagate through a medium since they continually fluctuate between
different configurations influencing particle dispersion. Thus properties of
the system become different from a situation in which all particles are
assumed to be completely independent of each other, behaving as freely
propagating bare particles.

Finite temperature background is particularly interesting as it offers more
realistic calculation and a possibility of measuring effects of the
polarized vacuum by employing thermodynamic quantities. When dealing with
sufficiently hot environments, the particles are considered to propagate in
background heat bath at energies around thresholds for particle antiparticle
pair production so that statistical effects due to temperature need to be
appropriately taken into account. These effects arise due to continuous
exchanges of mediating particles during physical interactions that take
place in a heat bath containing hot particles and antiparticles. Thermal
background effects are incorporated through radiative corrections and net
statistical effects enter the theory through fermion and boson
distributions. There is multitude of ways to describe physics at finite
temperature. In real-time formalism, manifest covariance is restored through 
$u^{\mu }=(1,0,0,0)$, the four component velocity of background heat bath.
Finite temperature calculations also provide a guideline to estimate density
corrections at higher order loops through chemical potential effects from
background plasma.

Quantum electrodynamics (QED) is useful to study the conditions for
existence of relativistic electron-positron plasmas. Self energies of
particles in high temperature medium relevant in QED, acquire temperature
corrections through virtual exchanges with real particles, in accordance
with perturbative nature of the theory. Mass less photons gain an effective
mass due to medium effects while propagating through a system comprising of
a cloud of electrons and positrons at finite temperature, and vice versa.
This gives rise to different terms in perturbation of states describing the
system. Many of these terms can be reproduced by modifying particle
propagators and from the poles of the propagators one obtains modified
dispersion relation. Radiatively generated mass shift acts as a kinematical
cut-off in production rate of particles in the heat bath. Gauge bosons
acquire a dynamically generated mass, at single and higher loop levels, due
to plasma screening effect. One loop corrections have been studied in detail
in real time formulation [32-35]. Medium properties such as electric
permittivity and magnetic susceptibility or permeability also get modified
by the thermal background effects.

The problem of renormalization in finite temperature theories is similar to
that at zero temperature. The temperature acts as a regularization parameter
for ultraviolet divergences. Infrared divergences introduced in finite
temperature framework are also appropriately removable in particle decay
processes, for example, via bremsstrahlung emission and absorption effects
[21,22]. Renormalization of QED in this scheme has been already checked in
detail at one loop level, using real time formulation, for all possible
ranges of temperatures and chemical potentials [20-28]. Method for
re-summation over hard thermal loops (HTL) is commonly used to determine the
higher loop corrections at finite temperature [29-31]. As far as
renormalization is concerned, as discussed in [31], HTL corrections are not
necessary.

We aim here to calculate photon self energy in a framework valid for all
relevant [32-49] temperatures in QED, that is below 5 MeV. To calculate the
photon self energy upto two loops, we are using the $1-1$ component of the
propagator, such that $Re\Sigma (p)=Re\Sigma _{11}(p)$ [33]. Doubling of the
fields is not necessary for the calculation of the real part of the self
energy that has been calculated in this work. The electron self-energy was
already calculated in detail, for all temperature ranges of relevance in QED
with renormalized mass, wave function and magnetic moment determined, at
two-loop level [41-48] in this scheme of calculations.

We derive an expression for renormalized QED coupling constant in a
background where electrons and positrons acquire temperature contribution
from the heat bath along with photons. These calculations have been done in
a manner that the limit of temperature $T\sim m,$ near the threshold for
creation of $e^{+}e^{-}$ pairs is inclusive. We now present calculations for
photon self energy in a general temperature framework, for the first time,
for renormalization of charge in this general framework of $T\sim m$ so that
the calculations are valid for all the temperature ranges relevant in QED
while the expressions in low temperature limit, $T<<$ $m$\ , presented in
[36], are reproducible from these results.{\small \ }

\section{\textbf{Two Loop Vacuum Polarization}$\qquad \qquad \qquad $}

In real time formalism, temperature dependent (hot) and temperature
independent (cold) terms are simply combined, at one loop level, because the
propagators include them separately as additive terms. At higher-loop level,
the loop integrals involve a combination of cold and hot terms which appear
due to the overlapping propagator terms. Therefore, the calculations are
much more involved.

\FRAME{ftbpFU}{3.2984in}{0.6858in}{0pt}{\Qcb{Vacuum polarization diagrams at
the second order in $\protect \alpha .$}}{}{fig1.eps}{\special{language
"Scientific Word";type "GRAPHIC";display "USEDEF";valid_file "F";width
3.2984in;height 0.6858in;depth 0pt;original-width 4.4131in;original-height
0.8622in;cropleft "0";croptop "1";cropright "1";cropbottom "0";filename
'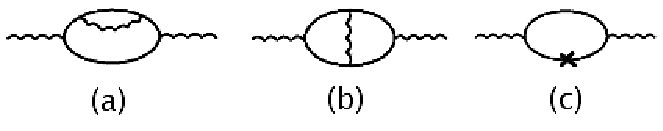';file-properties "XNPEU";}}

Vacuum polarization tensor of photon at the two-loop level contributes to
the second order hot corrections to charge renormalization constant of QED
at finite temperature. This contribution basically comes from self mass in
figure 1a and vertex type electron loop corrections inside the vacuum
polarization tensor in figure 1b. The expression for two loop self energy of
photon from self energy type correction in figure 1a is%
\begin{eqnarray}
\Pi _{\mu \nu }^{a}(p) &=&e^{4}\int \frac{d^{4}k}{(2\pi )^{4}}\int \frac{%
d^{4}q}{(2\pi )^{4}}  \notag \\
&&\times tr[\gamma _{\mu }S_{\beta }(k)\gamma _{\rho }D_{\beta }^{\rho
\sigma }(q-k)S_{\beta }(q)\gamma _{\sigma }S_{\beta }(k)\gamma _{\nu
}S_{\beta }(p-k)].
\end{eqnarray}%
with $k+l=q$. On substitution of the particle propagators at finite
temperature, this gives%
\begin{eqnarray}
\Pi _{\mu \nu }^{a}(p,T) &=&\Pi _{\mu \nu }^{a}(p,T=0)  \notag \\
&&+2\pi e^{4}\int \frac{d^{4}k}{(2\pi )^{4}}\int \frac{d^{4}q}{(2\pi )^{4}}%
N_{\mu \nu }^{a}[\frac{n_{F}(k)}{%
2E_{k}^{2}l^{2}[(p-k)^{2}-m^{2}](q^{2}-m^{2})}  \notag \\
&&\{ \frac{\delta ^{\prime }(k_{0}-E_{k})}{2}+\frac{\delta (k_{0}-E_{k})}{%
(k_{0}+E_{k})}\}-\frac{\delta \{q^{2}-m^{2}\}n_{F}(q)}{%
l^{2}[(p-k)^{2}-m^{2}](k^{2}-m^{2})^{2}}  \notag \\
&&-\frac{\delta \{(q-k)^{2}\}n_{B}(q-k)}{%
(k^{2}-m^{2})^{2}[(p-k)^{2}-m^{2}](q^{2}-m^{2})}  \notag \\
&&+\frac{\delta \{(p-k)^{2}-m^{2}\}n_{F}(p-k)}{%
(q-k)^{2}[k^{2}-m^{2}]^{2}(q^{2}-m^{2})}  \notag \\
&&+\frac{\delta \{(p-k)^{2}-m^{2}\}n_{F}(p-k)}{%
(q-k)^{2}(k^{2}-m^{2})^{2}(q^{2}-m^{2})}  \notag \\
&&-4\pi ^{2}\{ \frac{\delta \{q^{2}-m^{2}\}n_{F}(q)\delta
\{(p-k)^{2}-m^{2}\}n_{F}(p-k)}{(q^{2}-m^{2})^{2}}  \notag \\
&&\times \delta (q-k)^{2}n_{B}(q-k)[\frac{\delta ^{\prime }(k_{0}-E_{k})}{2}+%
\frac{\delta (k_{0}-E_{k})}{(k_{0}+E_{k})}]\frac{n_{F}(k)}{2E_{k}^{2}} 
\notag \\
&&+[\frac{\delta \{(p-k)^{2}-m^{2}\}n_{F}(p-k)\delta \{q^{2}-m^{2}\}n_{F}(q)%
}{(q-k)^{2}(k^{2}-m^{2})}  \notag \\
&&+\frac{\delta \{(q-k)^{2}\}n_{B}((q-k))\delta \{(p-k)^{2}-m^{2}\}n_{F}(p-k)%
}{(q^{2}-m^{2})(k^{2}-m^{2})}  \notag \\
&&+\frac{\delta \{(q-k)^{2}\}n_{B}(q-k)\delta \{q^{2}-m^{2}\}n_{F}(q)}{%
(k^{2}-m^{2})[(p-k)^{2}-m^{2}]}]\}],
\end{eqnarray}%
where

\begin{eqnarray}
N_{\mu \nu }^{a} &=&8[\left( 3m^{2}-k^{2}-2k.q\right) \{k_{\mu }(p-k)_{\nu
}+k_{\nu }(p-k)_{\mu }-g_{\mu \nu }k.(p-k)\}  \notag \\
&&+\left( k^{2}-m^{2}\right) \{(q-k)_{\mu }(p-k)_{\nu }+(q-k)_{\nu
}(p-k)_{\mu }  \notag \\
&&-g_{\mu \nu }(q-k).(p-k)\}+2g_{\mu \nu }m^{2}(m^{2}-k.q+k^{2})].
\end{eqnarray}%
and $\delta ^{\prime }(k_{0}-E_{k})$ is the derivative w.r.t. $k$.

Vertex type correction to two loop self energy of photon in figure 1b can be
written as

\begin{eqnarray}
\Pi _{\mu \nu }^{b}(p) &=&e^{4}\int \frac{d^{4}k}{(2\pi )^{4}}\int \frac{%
d^{4}q}{(2\pi )^{4}}  \notag \\
&&\times tr[\gamma _{\mu }S_{\beta }(k)\gamma _{\rho }D_{\beta }^{\rho
\sigma }(q-k)S_{\beta }(q)\gamma _{\nu }S_{\beta }(q-p)\gamma _{\sigma
}S_{\beta }(k-p)],
\end{eqnarray}%
\begin{eqnarray}
\Pi _{\mu \nu }^{b}(p,T) &=&\Pi _{\mu \nu }^{b}(p,T=0)+2\pi e^{4}\int \frac{%
d^{4}k}{(2\pi )^{4}}\int \frac{d^{4}q}{(2\pi )^{4}}N_{\mu \nu }^{b}  \notag
\\
&&\times \lbrack \{ \frac{1}{[(k-p)^{2}-m^{2}][(q-p)^{2}-m^{2}](q^{2}-m^{2})}
\notag \\
&&\times \lbrack \frac{\delta \{(q-k)^{2}\}n_{B}(q-k)}{(k^{2}-m^{2})}-\frac{%
\delta (k^{2}-m^{2})n_{F}(k)}{(q-k)^{2}}]  \notag \\
&&-\frac{1}{(q-k)^{2}[(k-p)^{2}-m^{2}](q^{2}-m^{2})}  \notag \\
&&\times \lbrack \frac{\delta (q^{2}-m^{2})n_{F}(q)}{(k^{2}-m^{2})}+\frac{%
\delta \{(q-p)^{2}-m^{2}\}n_{F}(q-p)}{(q^{2}-m^{2})}]  \notag \\
&&-\frac{\delta \{(k-p)^{2}-m^{2}\}n_{F}(k-p)}{%
(q-k)^{2}[(q-p)^{2}-m^{2}](k^{2}-m^{2})(q^{2}-m^{2})}\}  \notag \\
&&+4\pi ^{2}\{ \frac{\delta \{(q-k)^{2}\}n_{B}(q-k)\delta
(q^{2}-m^{2})n_{F}(q)}{[(k-p)^{2}-m^{2}]}  \notag \\
&&\times \lbrack \frac{\delta (k^{2}-m^{2})n_{F}(k)}{[(q-p)^{2}-m^{2}]}+%
\frac{\delta \{(q-p)^{2}-m^{2}\}n_{F}(q-p)}{(k^{2}-m^{2})}]  \notag \\
&&+\frac{\delta \{(q-k)^{2}\}n_{B}(q-k)\delta (k^{2}-m^{2})n_{F}(k)}{%
(q^{2}-m^{2})}  \notag \\
&&\times \lbrack \frac{\delta \{(k-p)^{2}-m^{2}\}n_{F}(k-p)}{%
[(q-p)^{2}-m^{2}]}  \notag \\
&&+\frac{\delta \{(q-p)^{2}-m^{2}\}n_{F}(q-p)}{[(k-p)^{2}-m^{2}]}]  \notag \\
&&-\frac{\delta (k^{2}-m^{2})n_{F}(k)\delta (q^{2}-m^{2})n_{F}(q)}{(q-k)^{2}}
\notag \\
&&\times \lbrack \frac{\delta \{(q-p)^{2}-m^{2}\}n_{F}(q-p))}{%
[(k-p)^{2}-m^{2}]}  \notag \\
&&-\frac{\delta \{(k-p)^{2}-m^{2}\}n_{F}(k-p)}{[(q-p)^{2}-m^{2}]}]  \notag \\
&&-\frac{\delta \{(q-p)^{2}-m^{2}\}n_{F}(q-p)\delta
\{(k-p)^{2}-m^{2}\}n_{F}(k-p)}{(q-k)^{2}}  \notag \\
&&\times \lbrack \frac{\delta (k^{2}-m^{2})n_{F}(k)}{(q^{2}-m^{2})}+\frac{%
\delta (q^{2}-m^{2})n_{F}(q)}{(k^{2}-m^{2})}]\}],
\end{eqnarray}%
where

\begin{eqnarray}
N_{\mu \nu }^{b} &=&-8[q.(k-p)\{k_{\mu }(q-p)_{\nu }-(q-p)_{\mu }k_{\nu }\} 
\notag \\
&&-q.(q-p)\{k_{\mu }(k-p)_{\nu }  \notag \\
&&+(k-p)_{\mu }k_{\nu }\}+(q-p).(k-p)\{k_{\mu }q_{\nu }+q_{\mu }k_{\nu }\} 
\notag \\
&&+k.q\{(k-p)_{\mu }(q-p)_{\nu }+(q-p)_{\mu \text{ }}(k-p)_{\nu }\}  \notag
\\
&&-k.(k-p)\{q_{\mu }(q-p)_{\nu }+(q-p)_{\mu }q_{\nu }\}  \notag \\
&&+k.(q-p)\{q_{\mu }(k-p)_{\nu }-(k-p)_{\mu }q_{\nu }\}  \notag \\
&&+g_{\mu \nu }\{q.(k-p)k.(q-p)-k.q(k-p).(q-p)  \notag \\
&&+k.(k-p)q.(q-p)\}-m^{2}\{2k_{\mu }(2q-p)_{\nu }\}  \notag \\
&&-[k_{\mu }(k-p)_{\nu }+(k-p)_{\mu }k_{\nu }+g_{\mu \nu }k.(p-k)]  \notag \\
&&-[q_{\mu }(q-p)_{\nu }+(q-p)_{\mu }q_{\nu }-g_{\mu \nu }q.(q-p)]  \notag \\
&&+2(k-p)_{\mu }(2q-p)_{\nu }\}+m^{4}g_{\mu \nu }].
\end{eqnarray}

We find that most of the terms vanish on integrating over hot momenta before
the cold ones. Therefore, we get respective nonzero contributions to these
diagrams, only from the terms:

\begin{eqnarray}
\Pi _{\mu \nu }^{a}(p,T &\neq &0)=-2\pi e^{4}\int \frac{d^{4}k}{(2\pi )^{4}}%
\int \frac{d^{4}q}{(2\pi )^{4}}N_{\mu \nu }^{a}  \notag \\
&&\frac{\delta (q-k)^{2}n_{B}(q-k)}{(k^{2}-m^{2})^{2}}[\frac{1}{%
\{(p-k)^{2}-m^{2}\}(q^{2}-m^{2})}  \notag \\
&&+4\pi ^{2}\delta (q^{2}-m^{2})n_{F}(q)\delta
\{(p-k)^{2}-m^{2}\}n_{F}(p-k)].
\end{eqnarray}%
and

\begin{eqnarray}
\Pi _{\mu \nu }^{b}(p,T &\neq &0)=8\pi ^{3}e^{4}\int \frac{d^{4}k}{(2\pi
)^{4}}\int \frac{d^{4}q}{(2\pi )^{4}}N_{\mu \nu }^{b}\text{ }\delta
(q-k)^{2}n_{B}(q-k)  \notag \\
&&\times \{ \frac{\delta (k^{2}-m^{2})n_{F}(k)\delta \lbrack
(q-p)^{2}-m^{2}]n_{F}(q-p)}{[(k-p)^{2}-m^{2}](q^{2}-m^{2})}\}.
\end{eqnarray}

Due to manifest covariance in the theory, physically measurable couplings
can be evaluated through contraction of vacuum polarization tensor $\Pi
_{\mu \nu }$ with the metric in Minkowski \ space $g^{\mu }{}^{\nu }$ and
the bath velocities $u^{\mu }u^{\nu }.$ Following [21], one can write

\begin{equation*}
\Pi _{\mu \nu }(p)=\Pi _{L}(p)P_{\mu \nu }+\Pi _{T}(p)Q_{\mu \nu },
\end{equation*}%
where 
\begin{equation*}
\Pi _{L}(p)=-\frac{p^{2}}{|\mathbf{p}|^{2}}u^{\mu }{}u^{\nu }\Pi _{\mu \nu },
\end{equation*}%
and 
\begin{equation*}
\Pi _{T}(p)=-\frac{1}{2}\Pi _{L}(p)+\frac{1}{2}g^{\mu }{}^{\nu }\Pi _{\mu
\nu }.
\end{equation*}%
The coefficients $P_{\mu \nu }$ and $Q_{\mu \nu }$\ are given by

\begin{equation*}
P_{\mu \nu }=(g_{\mu \nu }-u_{\mu }{}u_{\nu })+\frac{1}{|\mathbf{p}|^{2}}%
(p_{\mu }-\omega u_{\mu })(p_{\nu }-\omega u_{\nu }),
\end{equation*}%
and

\begin{equation*}
Q_{\mu \nu }=\frac{1}{p^{2}|\mathbf{p}|^{2}}[\{p^{2}u_{\mu }{}+\omega
(p_{\mu }-\omega u_{\mu })\} \{p^{2}u_{\nu }+\omega (p_{\nu }-\omega u_{\nu
})\}].
\end{equation*}
\newline

If a preferred order of integration (integrating over temperature dependent
variables before temperature independent ones) is not followed, the loop
calculations become very cumbersome and the results become messed up. In (7)
and (8), integrating over $d^{4}q$ (the temperature dependent momentum),
using Feynman parameterization and then integrating over $d^{4}k,$ after a
somewhat lengthy calculation, one gets

\begin{eqnarray}
u^{\mu }{}u^{\nu }\Pi _{\mu \nu }^{a+b} &=&2\alpha ^{2}[T^{2}\{ \frac{1}{3}(%
\frac{1}{2\eta }+1-\frac{E^{2}}{6m^{2}})  \notag \\
&&+4\dsum \limits_{_{n,r,s}=1}^{\infty }\frac{(-1)^{n+r}e^{-n\beta E}}{%
(n+s)(n-r)}e^{-m\beta (n-r)}\}  \notag \\
&&-\dsum \limits_{_{n,r,s}=1}^{\infty }(-1)^{s+r}e^{-s\beta E}\{ \frac{%
2m^{2}T}{(n+s)}[\{ \frac{e^{-m\beta (s+r)}}{m}  \notag \\
&&-\frac{(r+s)}{T}\func{Ei}[-m\beta (r+s)]  \notag \\
&&-\frac{T}{(r+s)}(4+\frac{5}{v}\ln \frac{1+v}{1-v})  \notag \\
&&-\frac{E}{v^{2}}\ln (\frac{1+v}{1-v})^{2}I_{A}\}]\}],
\end{eqnarray}%
whereas

\begin{eqnarray}
g^{\mu }{}^{\nu }\Pi _{\mu \nu }^{a+b} &=&-\alpha ^{2}[\frac{T^{2}}{3}(\frac{%
5}{\eta }-5\ln (-m^{2})-1+\frac{E^{2}}{m^{2}})  \notag \\
&&+48\dsum \limits_{_{n,r,s}=1}^{\infty }\{(-1)^{n+r+1}\frac{Te^{-n\beta E}}{%
(r-n)}[m+\frac{T}{(r-n)}\}  \notag \\
&&+e^{-s\beta E}(-1)^{s+r}\frac{T}{(n+s)}[\frac{1}{(r+s)}  \notag \\
&&-\frac{m^{2}}{2}\frac{e^{-m\beta (s+r)}}{m}-\frac{(r+s)}{T}\func{Ei}%
[-m\beta (r+s)]\}]\}],
\end{eqnarray}%
with $v=\frac{|\mathbf{p}|}{p_{_{0}}}$ and $\frac{1}{\eta }=\frac{1}{%
\varepsilon }-\gamma -\ln (\frac{4\pi \mu ^{2}}{m^{2}})$. \newline
We need to remove the temperature enhanced ultraviolet divergences. The
counter terms from figure 1c provide the contribution necessary to cancel
the divergences appearing in figures 1a and 1b. Adding the results from all
QED graphs in figures 1a - 1c, it has been explicitly checked that all the
hot and cold divergences cancel at the two loop level. Therefore, finite
terms from (9) and (10) are used to obtain longitudinal and transverse
components of the two loop vacuum polarization tensor. Including the one
loop corrections in [23] also, these components can be now written as

\begin{eqnarray}
\Pi _{L}(p,T) &=&-\frac{p^{2}}{|\mathbf{p}|^{2}}u^{\mu }u^{\nu }\Pi _{\mu
\nu }(p,T)  \notag \\
&=&\frac{16\alpha }{\pi }(1-\frac{1}{v^{2}})\{(1-\frac{1}{2v}\ln \frac{1+v}{%
1-v})(\frac{ma(m\beta )}{\beta }-\frac{c(m\beta )}{\beta ^{2}})  \notag \\
&&+\frac{b(m\beta )}{4}(2m^{2}-\omega ^{2}+\frac{11v^{2}+37}{72}E^{2})\} 
\notag \\
&&-\frac{2\alpha ^{2}}{v^{2}}[\frac{T^{2}}{3}(1+\frac{E^{2}}{2m^{2}}%
)+4T^{2}\dsum \limits_{_{n,r,s}=1}^{\infty }e^{-n\beta E}\frac{%
(-1)^{n+r}e^{-m\beta (n-r)}}{(n+s)(n-r)}  \notag \\
&&+\dsum \limits_{_{n,r,s}=1}^{\infty }(-1)^{s+r}e^{-s\beta E}\{ \frac{T}{%
(n+s)}[\frac{T}{(r+s)}(4+\frac{5}{v}\ln \frac{1+v}{1-v})  \notag \\
&&-2m^{2}\{ \frac{e^{-m\beta (s+r)}}{m}-(r+s)\beta \func{Ei}[-m\beta
(r+s)]\}]\}],
\end{eqnarray}

and 
\begin{eqnarray}
\Pi _{T}(p,T) &=&-\frac{1}{2}[\Pi _{L}(p,T)-g^{\mu }{}^{\nu }\Pi _{\mu \nu
}(p,T)]  \notag \\
&=&\frac{8\alpha }{\pi }[\{[\frac{1}{v^{2}}+(1-\frac{1}{v^{2}})\ln \frac{1+v%
}{1-v}\}(\frac{ma(m\beta )}{\beta }-\frac{c(m\beta )}{\beta ^{2}})  \notag \\
&&+\frac{1}{8}(2m^{2}+E^{2}[1+\frac{107-131v^{2}}{72}])b(m\beta )]  \notag \\
&&+\alpha ^{2}[\frac{T^{2}}{3}\{ \frac{1}{2}-\frac{1}{v^{2}}(1+\frac{E^{2}}{%
2m^{2}})  \notag \\
&&+\dsum \limits_{_{n,r,s}=1}^{\infty }(-1)^{s+r}e^{-s\beta E}\frac{T}{(n+s)}%
\{T[\  \frac{4}{v^{2}}+\frac{24}{(r+s)}  \notag \\
&&+\frac{5}{(r+s)v}\ln \frac{1+v}{1-v}-2m^{2}(\frac{1}{v^{2}}+6)[\frac{%
e^{-m\beta (s+r)}}{m}  \notag \\
&&-(r+s)\beta \func{Ei}\{-m\beta (r+s)\}]\}].
\end{eqnarray}

where

\begin{equation}
a(m\beta )=\ln (1+e^{-m\beta }),
\end{equation}

\begin{equation}
b(m\beta )=\dsum \limits_{_{n=1}}^{\infty }(-1)^{n}\func{Ei}(-nm\beta ),
\end{equation}%
\begin{equation}
c(m\beta )=\dsum \limits_{_{n=1}}^{\infty }(-1)^{n}\frac{e^{-nm\beta }}{n^{2}%
},
\end{equation}%
$I_{A}$ is infrared divergence at finite temperature and $\func{Ei}$ is
error integral given by

\begin{equation}
\func{Ei}(-x)=-\int_{x}^{\infty }\frac{dt}{t}e^{-t}.
\end{equation}%
These expressions are in a form such that all the relevant limits of
temperature including $T\sim m$ are valid.

\section{\textbf{Charge Renormalization in QED at High Temperature}$\qquad
\qquad $}

Virtual photons are emitted and reabsorbed by electrons and tend to smear
out charge distribution and correct the electron mass. The virtual fermion
loops get polarized in electromagnetic fields, which act to damp the field.
Renormalized electron mass and renormalized wavefunction have been already
calculated in detail [32]. Vacuum polarization in a medium gives
modification to electric charge and QED coupling constant. The electric
charge couples with the medium through vacuum polarization get screened, and
picks up thermal corrections accordingly. We derive an expression for the
renormalized charge following [21] and using (11). The electron charge
renormalization constant up to the order $\alpha ^{2}$ can therefore be
expressed as:

\begin{eqnarray}
Z_{3} &=&1-\frac{\alpha }{3\pi \varepsilon }  \notag \\
&&+\frac{8\alpha }{\pi m^{2}}[\frac{ma(m\beta )}{\beta }-\frac{c(m\beta )}{%
\beta ^{2}}+\frac{b(m\beta )}{4}(m^{2}+\frac{1}{3}\omega ^{2})]  \notag \\
&&+\frac{\alpha ^{2}}{m^{2}}[\frac{T^{2}}{6}+\dsum
\limits_{_{n,r,s}=1}^{\infty }(-1)^{s+r}e^{-s\beta E}\frac{T}{(n+s)}\{ \frac{%
24T}{(r+s)}  \notag \\
&&-12m^{2}[\frac{e^{-m\beta (s+r)}}{m}-(r+s)\beta \func{Ei}\{-m\beta
(r+s)\}]\}].\qquad
\end{eqnarray}%
Using $Z_{3}$, a corresponding expression for the renormalized QED\ coupling
constant is obtained to be:

\begin{eqnarray}
\alpha _{R} &=&\{ \alpha (T=0)  \notag \\
&&+\frac{8\alpha }{\pi m^{2}}[\frac{ma(m\beta )}{\beta }-\frac{c(m\beta )}{%
\beta ^{2}}+\frac{b(m\beta )}{4}(m^{2}+\frac{1}{3}\omega ^{2})]  \notag \\
&&+\frac{\alpha ^{2}}{m^{2}}[\frac{T^{2}}{6}+\dsum
\limits_{_{n,r,s}=1}^{\infty }(-1)^{s+r}e^{-s\beta E}\frac{T}{(n+s)}\{ \frac{%
24T}{(r+s)}  \notag \\
&&-12m^{2}[\frac{e^{-m\beta (s+r)}}{m}-(r+s)\beta \func{Ei}\{-m\beta
(r+s)\}]\}]\}.\qquad
\end{eqnarray}

It can be seen from the above equations that electron charge and hence the
QED coupling constant gets smaller with an increased order of loops which
clearly assures the renormalization of electron charge. This modification in
the coupling constant can be used to determine changes in electromagnetic
properties of medium. The leading two loop corrections to $Z_{3}$ and $%
\alpha _{R}$\ come out to be $\frac{\alpha ^{2}}{6}\approx 8.88\times
10^{-6} $ at $T\sim m$. The values of the renormalization constant $Z_{3}$\
are plotted vs $T$\ in figure 2. A comparison between the one-loop and
two-loop contribution is presented in figure 2 and is interesting to
analyse. It can be seen from figure 2 that the major contributions at $T\geq
m$\ come from the second order corrections. However, at $T$ sufficiently
larger than $m$, there is a crossover between the one-loop and two-loop
corrections with the second order contributions become more significant than
the first order contributions around temperatures when $T\sim m$. Then the
second order contributions exponentially increase beyond this temperature
whereas the one loop contributions go up slowly. Since the second order
contributions approach unity much rapidly, the higher order corrections may
be expected to rise even faster than those for two-loop level. This may be
due to problems with perturbative QED renormalization at finite temperature
at higher orders.

\FRAME{ftbpFU}{3.5172in}{2.4241in}{0pt}{\Qcb{Comparison of first order and
second order contributions in all the temperature ranges for $Z_{3}$.}}{}{%
fig2.eps}{\special{language "Scientific Word";type
"GRAPHIC";maintain-aspect-ratio TRUE;display "USEDEF";valid_file "F";width
3.5172in;height 2.4241in;depth 0pt;original-width 3.4714in;original-height
2.3834in;cropleft "0";croptop "1";cropright "1";cropbottom "0";filename
'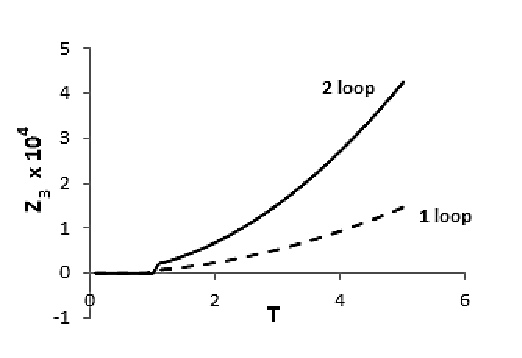';file-properties "XNPEU";}}

\section{Electromagnetic Properties of the Medium}

It is a fact that the electromagnetic properties of media depend on their
statistical properties. Higher order calculation of thermal corrections to
self energy of photons is much more complicated. However, it is interesting
to determine the corresponding changes in electromagnetic properties of hot
media which arise due to radiative emission and absorption of hot particles
in the heat bath. It can be checked using [22] that electric permittivity of
a medium at the two loop level modifies to

\begin{eqnarray}
\varepsilon (p,T) &=&1-|\mathbf{p}|^{2}\Pi _{L}(p,T)  \notag \\
&=&1-\frac{16\alpha m^{2}}{\pi v^{2}E^{2}}\{(1-\frac{1}{2v}\ln \frac{1+v}{1-v%
})(\frac{ma(m\beta )}{\beta }-\frac{c(m\beta )}{\beta ^{2}})  \notag \\
&&+\frac{b(m\beta )}{4}[2m^{2}-E^{2}(1-\frac{11v^{2}+37}{72})]\}  \notag \\
&&+2\alpha ^{2}p^{2}[\frac{T^{2}}{3}(1+\frac{E^{2}}{2m^{2}})-4T^{2}\dsum
\limits_{_{n,r,s}=1}^{\infty }e^{-n\beta E}\frac{(-1)^{n+r}e^{-m\beta (n-r)}%
}{(n+s)(r-n)}  \notag \\
&&+\dsum \limits_{_{n,r,s}=1}^{\infty }(-1)^{s+r}e^{-s\beta E}\{ \frac{T}{%
(n+s)}[\frac{T}{(r+s)}(4+\frac{5}{v}\ln \frac{1+v}{1-v})  \notag \\
&&-2m^{2}\{ \frac{e^{-m\beta (s+r)}}{m}-(r+s)\beta \func{Ei}[-m\beta
(r+s)]\}]\}],
\end{eqnarray}%
whereas magnetic permeability becomes

\begin{eqnarray}
\frac{1}{\mu (p,T)} &=&1+\frac{1}{p^{2}}[\Pi _{T}(p,T)-\frac{1}{v^{2}}\Pi
_{L}(p,T)]  \notag \\
&=&1+\frac{8\alpha }{\pi v^{2}m^{2}}[\frac{1}{8}\{E^{2}(1+\frac{109v^{2}-129%
}{72})-6m^{2}\}b(m\beta )  \notag \\
&&+(\frac{1}{v^{2}}-1)(1-\frac{1-v^{2}}{2v}\ln \frac{1+v}{1-v})(\frac{%
ma(m\beta )}{\beta }-\frac{c(m\beta )}{\beta ^{2}})]  \notag \\
&&+\alpha ^{2}[\frac{T^{2}}{3}\{ \frac{1}{2m^{2}}-\frac{1}{E^{2}v^{2}}(1+%
\frac{E^{2}}{2m^{2}})(1+\frac{2}{v^{2}})\}  \notag \\
&&+\dsum \limits_{_{n,r,s}}(-1)^{s+r}e^{-s\beta E}\frac{T}{(n+s)}\{ \frac{4T%
}{(s-r)}e^{-m\beta (s-r)}(1+\frac{1}{v^{2}})\}  \notag \\
&&+\frac{T}{(r+s)}(4+\frac{5}{v}\ln \frac{1+v}{1-v})-2m^{2}(6+\frac{1}{v^{2}}%
)  \notag \\
&&\times \{ \frac{e^{-m\beta (s+r)}}{m}-(r+s)\beta \func{Ei}[-m\beta
(r+s)]\}]\}].
\end{eqnarray}%
Equations (19) and (20) show that the radiative emissions and absorptions of
real hot photons from the heat bath up to order $\alpha ^{2}$\ lead to
deviations from unity for the values of dielectric constant and magnetic
permeability. There are two possible limits for $\Pi _{L}$ and $\  \Pi _{T}$
[22] as $p^{2}\longrightarrow 0$ and their physical interpretations depend
on the order of taking the limits $p_{0}=0$ and $|\mathbf{p}|\longrightarrow
0$. In (11) and (12), if we take $p_{0}=$ $|\mathbf{p}|$ in the rest frame
of the heat bath and then limit $|\mathbf{p}|\longrightarrow 0$ is taken, we
get \bigskip 
\begin{eqnarray}
&&\kappa _{L}^{2}\  \text{\  \ }_{\widetilde{}}\  \lim_{|\mathbf{p}%
|\longrightarrow 0}\Pi _{L}(0,|\mathbf{p}|,T)  \notag \\
&\approx &\frac{16\alpha }{\pi }\{ \frac{ma(m\beta )}{\beta }-\frac{c(m\beta
)}{\beta ^{2}}+\frac{m^{2}b(m\beta )}{2}\}  \notag \\
&&+2\alpha ^{2}\{[\{ \frac{T^{2}}{3}-4T^{2}\dsum
\limits_{_{n,r,s}=1}^{\infty }\frac{(-1)^{n+r}}{(n+s)(r-n)}e^{-m\beta (n+r)}
\notag \\
&&+\dsum \limits_{_{n,r,s}=1}^{\infty }(-1)^{s+r}[\frac{4T^{2}}{(n+s)(r+s)} 
\notag \\
&&-2m^{2}\{ \frac{e^{-m\beta (s+r)}}{m}-(r+s)\beta \func{Ei}[-m\beta
(r+s)]\}]\},
\end{eqnarray}%
$\qquad \qquad $\ 
\begin{eqnarray}
&&\kappa _{T}^{2}\  \text{\  \ }_{\widetilde{}}\  \lim_{|\mathbf{p}%
|\longrightarrow 0}\Pi _{T}(0,|\mathbf{p}|,T)  \notag \\
&\approx &\frac{\alpha }{\pi }m^{2}b(m\beta )-\alpha ^{2}[\frac{T^{2}}{3} 
\notag \\
&&+\dsum \limits_{_{n,r,s}=1}^{\infty }(-1)^{s+r}\frac{T}{(n+s)}\{4T\frac{%
e^{-m\beta (s-r)}}{s-r}  \notag \\
&&-10m^{2}[\frac{e^{-m\beta (s+r)}}{m}-(r+s)\beta \func{Ei}\left \{ -m\beta
(r+s)\right \} ]  \notag \\
&&-\frac{24T}{(r+s)}+6(m+\frac{T}{r-s})\}],
\end{eqnarray}%
On the other hand if we set $p_{0}=0$ with $|\mathbf{p}|\longrightarrow 0$
then this results in:

\begin{equation}
\omega _{L}^{2}\text{\  \ }_{\widetilde{}}\  \lim_{|\mathbf{p}|\longrightarrow
0}\Pi _{L}(|\mathbf{p}|,|\mathbf{p}|,T)=0,
\end{equation}

\begin{eqnarray}
&&\omega _{T}^{2}\text{\  \ }_{\widetilde{}}\  \  \lim_{|\mathbf{p}%
|\longrightarrow 0}\Pi _{T}(|\mathbf{p}|,|\mathbf{p}|,T)  \notag \\
&\approx &\frac{8\alpha }{\pi }\{ \frac{ma(m\beta )}{\beta }-\frac{c(m\beta )%
}{\beta ^{2}}+\frac{m^{2}b(m\beta )}{4}\}  \notag \\
&&+\alpha ^{2}[\frac{T^{2}}{6}+\dsum \limits_{_{n,r,s}=1}^{\infty }(-1)^{s+r}%
\frac{T}{(n+s)}\{ \frac{24T}{(r+s)}  \notag \\
&&-12m^{2}[\frac{e^{-m\beta (s+r)}}{m}-(r+s)\beta \func{Ei}\{-m\beta (r+s)\}]
\notag \\
&&+24T(\frac{e^{-m\beta (s-r)}}{(s-r)}m+\frac{T}{r-s})\}].
\end{eqnarray}

The vacuum polarization tensor in $p^{2}\longrightarrow 0$ limit provides
the dynamically generated mass of photon.

\section{Results and Discussion}

Thermal corrections at the two loop level are not as simple as the one loop.
Temperature dependence to vacuum polarization tensor at the two loop level
enter as an overlap of photon and fermions hot loop with momenta due to
effectively mutual interactions while propagating through the medium. An
unusual behavior of hot integrals appears due to the overlap of hot and cold
terms. The order of integration between the cold and hot loop not only
affect the length of calculations but also the right order of integration is
required to show the renormalizability of the theory. With the integration
over temperature dependent variables before temperature independent ones,
the loop calculations become lesser cumbersome and also the cancellation of
singularities can be clearly seen. In case of wavefunction, the terms with
the reverse order of integrations have to be included to fully establish the
renormalization. The standard regularization techniques of vacuum theory
like dimensional regularization would only be valid in a covariant framework
of a Lorentz invariant system. Once the hot momenta are integrated out,
usual field theoretical techniques of Feynman parameterization and
dimensional regularization can be applied to remove singularities arising
from cold loops. The results reduce to finite values after order by order
cancellation of singularities on addition of counter terms of the same order.

At the higher loop level, temperature corrections in hot medium imply
convergence of the perturbative series. The renormalized self energies are
used to obtain an estimation of dielectric constant and magnetic
permeability of a hot medium. Both of these quantities deviate from unity
even for $T<<m$ whereas they do not deviate from the vacuum value with the
first order loop corrections [23]. This difference of loop correction can be
interpreted as at the one loop level self-interaction of the photon loops
does not contribute whereas at the two loop level the hot photon loop
develops self-interaction due to the dynamically generated mass. This type
of effect has been earlier observed for self-mass of gluon even at the
one-loop level [37] due to self-coupling of gluons. With an increase in
temperature, photon mass sufficiently affects the behavior of this coupling,
especially when $T\gtrsim m$. This is due to the self mass of photon at
finite temperature which may lead to nonzero self interaction.

Further we obtain respective propagation vectors and frequencies when we
take the limiting values for longitudinal and transverse component of vacuum
polarization tensor in (7) and (8). In particular from (21) $\kappa
_{L}^{-1} $\ gives Debye screening length of electric force in QED plasma
whereas the transverse component of vacuum polarization tensor in eq. (12)
corresponds to the dynamically generated mass of photon. QED effective
coupling is determined in (18).

The framework of calculations adopted here has a form such that the limits
of temperature of physical interest classified as, high temperature $(T>m,$
with $e^{-m\beta }$ ignorable as compared to $\frac{T^{2}}{m^{2}})$,
intermediate temperatures $T\sim m$ (by taking $m\beta \sim 1$) and low
temperature $T<<m$ (with fermions contribution negligible) are retrievable
from the results obtained in previous sections. From (17) and (18), the
leading two loop corrections to the renormalized coupling $\alpha _{R}$\
come out to be $\frac{\alpha ^{2}}{6}\approx 8.88\times 10^{-6}$ at $T\sim m$%
.

Since the temperature dependence for $T\geq m$ is contributed by the fermion
background, and the fermions are added to the system when the temperature
rises beyond the electron mass. This contribution is negligible below the
electron mass both at the one loop level as well as at the two loop level.
We use the calculations of vacuum polarization to determine the dynamically
generated mass of photon, Debye screening length, plasma frequency up to
order $\alpha ^{2}$ as well as the electromagnetic properties of a medium at 
$m\leq T\leq 2m$ temperature. For higher temperatures, QED coupling constant
keeps on increasing with temperature and the existing renormalization scheme
[41-50] is not useful. To exactly determine the validity of renormalization
scheme, even the higher order calculations are required.  The temperature $%
T\sim m$ is of specific interest from the point of view of the early
universe and stellar systems. Such calculations have acquired more
significance recently due to the possibility of producing $e^{+}e^{-}$
plasmas in laboratory .

It is worth-mentioning here that all hot corrections give a dominant $T^{2}$
dependence, as in case of low temperature. Two loop calculations with
imaginary time formulation were done earlier for hard lepton pair production
in the context of quark gluon plasma .

\textbf{References}{\small \ }

\begin{enumerate}
\item T. Matsubara, \textit{Prog. Theor. Phys.} \textbf{14} (1955) 351.

\item D. A. Kirzhnits,\ Field Theoretical Methods in Many Body Systems
(Permagon, Oxford, 1967).

\item E. M. Lifshitz and L. P. Pitaevskii, Course on Theoretical Physics-
Physical Kinetics (Pergamon Press, New York, 1981).

\item J. Schwinger, \textit{J. Math. Phys.} \textbf{2} (1961) 407.

\item L. V. Keldysh, \textit{Sov. Phys.} \textbf{20} (1964) 1018.

\item Samina Masood, BAPS.2013.APR.S2.11: \textbf{arXiv:1205.2937 [hep-ph]}.

\item H. Umezawa, H. Matsumoto and M. Tachiki, Thermo Field Dynamics and
Condensed States (North Holland, Amsterdam, 1982); .

\item R. Mills, Propagators for Many Particle Systems (Gordon and Breach,
New York, 1969).

\item L. P. Kadanoff and G. Baym, Quantum Statistical Mechanics, (W. A.
Benjamin, Reading, 1978).

\item A. A. Abrikosov, L. P. Gorkov, and I. E. Dzyaloshinski, Methods of
Quantum Field Theory in Statistical Physics (Dover, New York, 1975).

\item A. Fetter and J.D. Walecka, Quantum Theory of Many Particle Systems
(Mc Graw Hill, New York, 1971).

\item J. I. Kapusta and C. Gale, Finite Temperature Field Theory (Cambridge
University Press, New York, 2006).

\item B. Muller, Physics of the Quark-Gluon Plasma, Lecture Notes in Physics
225 (Springer, Berlin, 1985).

\item C.-Y. Wong,\ Introduction to High-Energy Heavy-Ion Collisions (World
Scientific Pub Co, 1994).

\item J. Bartke, Relativistic Heavy Ion Physics (World Scientific,
Singapore, 2003).

\item M. LeBellac, Thermal Field Theory (Cambridge University Press, 1996).

\item A. Das, Finite Temperature Field Theory (World Scientific, Singapore,
1997).

\item S. Weinberg, Gravitation and Cosmology (Wiley, New York, 1972).

\item V. Mukhanov, Physical Foundations of Cosmology (Cambridge University
Press, 2005).

\item W. Dittrich and H. Gies, Probing the quantum vacuum (Springer-Verlag,
Berlin, Germany, 2000).

\item P. H. Cox, W. S. Hellman, and A. Yildiz, \textit{Ann. Phys.} \textbf{%
154} (1984) 211.

\item M. Loewe and J. C. Rojas, \textit{Phys. Rev. }\textbf{D 46} (1992)
2689.

\item P. Elmfors and B-S. Skagerstam, \textit{Phys. Lett.} \textbf{B 348}
(1995) 141.

\item N. B. Narozhny et al., \textit{Phys. Lett.} \textbf{A 330} (2004) 1.

\item A. R. Bell and John G. Kirk, \textit{Phys. Rev. Lett.} \textbf{101}
(2008) 200403.

\item J. G. Kirk, A. R. Bell, and I. Arka, \textit{Plasma Phys. Control.
Fusion} \textbf{51} (2009) 085008.

\item E.P. Liang, S.C. Wilks, and M. Tabak, \textit{Phys. Rev. Lett.} 
\textbf{81} (1998) 4887.

\item B. Shen, \textit{Phys. Rev.} \textbf{E 65} (2001) 016405.

\item M.H. Thoma, \textit{Rev. Mod. Phys.} \textbf{81} (2009) 959.

\item Ben King, Antonino Di Piazza, and Christoph H. Keitel, \textit{Nature
Photonics} \textbf{4} (2010) 92.

\item V. Yanovsky et al., \textit{Opt. Express} \textbf{16} (2008) 2109.

\item H. A. Weldon, \textit{Phys. Rev.} \textbf{D 26} (1982) 1394.

\item J. F. Donoghue, B. R. Holstein, and R. W. Robinett, \textit{Ann. Phys.
(N.Y.) }\textbf{164} (1985) 233.

\item K. Ahmed and Samina S. Masood, \textit{Ann. Phys. (N.Y.) } \textbf{207}
(1991) 460.

\item J. F. Donoghue and B. R. Holstein, \textit{Phys. Rev.} \textbf{D 28}
(1983) 340; \textit{ibid} \textbf{29} (1983) 3004(E).

\item Samina S. Masood, \textit{Phys. Rev.} \textbf{D 44} (1991) 3943.

\item Samina S. Masood, \textit{Phys. Rev.} \textbf{D 47} (1993) 648.

\item W. Keil and R. Kobes, \textit{Physica} \textbf{A 158} (1989) 47.

\item K. Ahmed and Samina Saleem, \textit{Phys. Rev.} \textbf{D 35} (1987)
1861.

\item K. Ahmed and Samina Saleem, \textit{Phys. Rev.} \textbf{D 35} (1987)
4020.

\item Samina S\ Masood,`Nucleosynthesis in Hot and Dense Media-, ( submitted
for publication ).

\item Mahnaz Qader, Samina S. Masood, and K. Ahmed, \textit{Phys. Rev.} 
\textbf{D 44} (1991) 3322.

\item Mahnaz Qader, Samina S. Masood, and K. Ahmed, \textit{Phys. Rev.} 
\textbf{D 44} (1991) 5633.

\item Mahnaz Q. Haseeb and Samina S. Masood, \textit{Chin. Phys.} \textbf{C
35} (2011) 608.

\item Mahnaz Q. Haseeb and Samina S. Masood, \textit{Phys. Lett.} \textbf{B
704} (2011) 66.

\item Samina S. Masood and Mahnaz Q. Haseeb, \textit{Int. J. of Mod. Phys.} 
\textbf{A 27} (2012) 1250188.

\item Samina S. Masood and Mahnaz Q. Haseeb, \textit{Int. J. of Mod. Phys.} 
\textbf{A 23} (2008) 4709.

\item Samina S. Masood and Mahnaz Q. Haseeb, Astropart. Phys. \textbf{3}
(1995) 405.

\item Samina S. Masood and Mahnaz Qader, \textit{Phys. Rev.} \textbf{D 46}
(1991) 5110.

\item Samina S. Masood and Mahnaz Qader, Finite temperature and density
corrections to electroweak decays, in Proc. 4th Regional Conference on
Mathematical Physics, Sharif Univ. of Technology, Tehran, Iran, 12--17 May
1990, eds. F. Ardalan, H. Arafae and S. Rouhani (Sharif University of
Technology Press, 1990), p. 334.
\end{enumerate}

Acknowledgement

One of the authors (MQH) thanks Higher Education Commission, Pakistan for
partial funding under a research grant no. 1925 during this work.

Figure caption

Figure 1. Vacuum polarization diagrams at the second order in $\alpha .$

Figure 2. Comparison of first order and second order contributions in all
the temperature ranges for $Z_{3}$.

\end{document}